\documentclass[
aps,prd,
12pt,%10pt
nopreprintnumbers,
showpacs,
eqsecnum,
nofootinbib
]{revtex4-1}

\usepackage{graphicx}
\usepackage{amssymb}

\def\tr{{\rm tr\,}}

\begin{document}

\title{GR--GSG Hybrid Gravity}
\author{Nahomi Kan}\email[]{kan@gifu-nct.ac.jp}
\affiliation{National Institute of Technology, Gifu College,
Motosu-shi, Gifu 501-0495, Japan
}
\author{Takuya Maki}\email[]{maki@jwcpe.ac.jp}
\affiliation{Japan Women's College of Physical Education,
Setagaya, Tokyo 157-8565, Japan
}
\author{Kiyoshi Shiraishi}\email[]{shiraish@yamaguchi-u.ac.jp}
\affiliation{
Graduate School of Sciences and Technology for Innovation, Yamaguchi
University, Yamaguchi-shi, Yamaguchi 753--8512, Japan}
\date{\today}
%\date{}

\begin{abstract}
We propose a model of gravity in which a General Relativity metric tensor
and an effective metric generated from a single scalar formulated in
Geometric Scalar Gravity are mixed. We show that the model yields the
exact Schwarzschild solution, along with accelerating behavior of
scale factors in cosmological solutions.
\end{abstract}

%\preprint{}


\maketitle

%%%%%%%%%%%%%%%%%%%%%%%%%%%%%%%%%%%%%%%%%%%%%%%%%%%%%%%%%%%%%%%%%%%%%%%%%%%
%Introduction
%%%%%%%%%%%%%%%%%%%%%%%%%%%%%%%%%%%%%%%%%%%%%%%%%%%%%%%%%%%%%%%%%%%%%%%%%%%
%%%%%%%%%%%%%%%%%%%%%%%%%%%%%%%%%%%%%%%%%%%%%%%%%%%%%%%%%%%%%%%%%%%%%%%%%%%
\section{Introduction}
\label{sec1}
%%%%%%%%%%%%%%%%%%%%%%%%%%%%%%%%%%%%%%%%%%%%%%%%%%%%%%%%%%%%%%%%%%%%%%%%%%%

Bigravity theory is an alternative theory of gravity
that describes both massless and massive gravitons \cite{HR1,HR2} (for
a review, see \cite{SS}).
The low-energy modification from General Relativity (GR) is expected to
explain mysterious components of the universe known as
dark matter and dark energy. 
Bigravity models generally contain two Einstein--Hilbert
terms representing the two metrics, as well as
the mixing term of the two metrics, in order to generate mass for a
graviton. Recent developments in the study of bigravity or massive gravity
theories \cite{RGT,HR,HRS,Hin,deR} have stemmed from the discovery of the
appropriate mass term for ghost-free nonlinear bimetric action.%
\footnote{In a certain sense, massive gravity is just bigravity in
which one of the metrics is non-dynamic.}

More recently, Nojiri, Odintsov, and their collaborators have considered
 extensions of the bigravity theory \cite{NO,NOS,BMMNO,BNO,NOO,BKNS}
in which the pure Einstein--Hilbert terms are replaced by the
Lagrangian of $F(R)$ gravity or scalar--tensor theories.%
\footnote{For the various models of modified gravity
and their cosmological meanings, see \cite{MG1,MG2,MG3}.}
These authors have studied models having scalar degrees of
freedom with a view to resolving the cosmological problems in the very
early era, as well as in the present universe.

In 2014, Novello and collaborators presented a new
theory of gravity, called Geometric Scalar Gravity (GSG) 
\cite{NBMGST,NB,BMNT1,BNMGST,Thesis,BMNT2}.%
\footnote{See also \cite{JL,KS}.} 
In this theory, the dynamics of gravity is described by a single scalar
field. A normalized derivative of the scalar field expresses part
of the dynamical metric, as well as the scalar field itself. 
Novello and his collaborators found a specific form of the scalar
field potential from which the Schwarzschild spacetime is derived
as an exact solution. These researchers also discussed the (exotic)
cosmology based on GSG
\cite{NBMGST,NB,BMNT1,BNMGST,Thesis,BMNT2}. 
The novel behavior of the scale factor in the GSG cosmology provides a 
very interesting supplementary perspective on the issue of the
initial singularity \cite{BNMGST}. On the other hand, GSG predicts scalar
gravitational waves \cite{NBMGST,TN}, which may conflict with
recent direct observations of gravitational waves from a black hole
binary \cite{LIGO}.
Further, more scalar degrees of
freedom may be needed in order to explain the gravitational field around
a spinning source \cite{BMNT2,Novello}.
Therefore, the simplest GSG model has practical difficulty in describing
astrophysical processes.

In this paper, we propose a GR--GSG hybrid model of gravity.
Our model consists of the dynamics of a fundamental metric tensor in GR
and an effective metric in GSG.  This theory naturally possesses a
massless mode for the symmetric tensor field and yields the
Schwarzschild solution exactly. Nevertheless, the cosmological solutions
in the model are expected to be interesting, because they may inherit
novel characteristics from GSG solutions. In the next section, we
define our model.  In Sec.~\ref{sec3}, we investigate a spherically
symmetric solution of the model in  weak gravity.  In
Sec.~\ref{sec4},  we explore cosmological solutions for our model. We
consider two cases: two metrics independently
coupled to corresponding matter and the case in which the
``composite'' metric is considered to be the physical metric coupled to
matter. Finally, we summarize
our work and remark on the general significance of our study in
Sec.~\ref{summary}.

%%%%%%%%%%%%%%%%%%%%%%%%%%%%%%%%%%%%%%%%%%%%%%%%%%%%%%%%%%%%%%%%%%%%%%%%%%%
%%%%%%%%%%%%%%%%%%%%%%%%%%%%%%%%%%%%%%%%%%%%%%%%%%%%%%%%%%%%%%%%%%%%%%%%%%%
\section{The GR--GSG hybrid model}
\label{sec2}
%%%%%%%%%%%%%%%%%%%%%%%%%%%%%%%%%%%%%%%%%%%%%%%%%%%%%%%%%%%%%%%%%%%%%%%%%%%
%%%%%%%%%%%%%%%%%%%%%%%%%%%%%%%%%%%%%%%%%%%%%%%%%%%%%%%%%%%%%%%%%%%%%%%%%%%

%%%%%%%%%%%%%%%%%%%%%%%%%%%%%%%%%%%%%%%%%%%%%%%%%%%%%%%%%%%%%%%%%%%%%%%%%%%
\subsection{Brief review of GSG}
%%%%%%%%%%%%%%%%%%%%%%%%%%%%%%%%%%%%%%%%%%%%%%%%%%%%%%%%%%%%%%%%%%%%%%%%%%%
First, we provide a brief review  of GSG
\cite{NBMGST} to render the present paper self-contained.
The effective metric $q_{\mu\nu}$ in GSG is described by a scalar field
$\Phi$ as
\begin{equation}
q_{\mu\nu}=e^{2\Phi}\left[\eta_{\mu\nu}-
\frac{e^{-4\Phi}V(\Phi)-1}{e^{-4\Phi}V(\Phi)}\frac{\partial_\mu\Phi\partial_\nu\Phi}{w}
\right]\,,
\end{equation}
where
$w\equiv\eta^{\mu\nu}\partial_\mu\Phi\partial_\nu\Phi$ and
$\eta_{\mu\nu}$ is a flat Minkowski metric with the signature
$(-+++)$.
The inverse of the effective metric is then written as
\begin{equation}
q^{\mu\nu}=e^{-2\Phi}\left[\eta^{\mu\nu}+\frac{e^{-4\Phi}V(\Phi)-1}{w}
\eta^{\mu\rho}\eta^{\nu\sigma}\partial_\rho\Phi\partial_\sigma\Phi\right]
\,.
\end{equation}
Note that
\begin{equation}
\sqrt{-\det
q}=\frac{e^{6\Phi}}{\sqrt{V(\Phi)}}\sqrt{-\det\eta}\,,\quad
q^{\mu\nu}\partial_\nu\Phi=e^{-6\Phi}V(\Phi)\eta^{\mu\nu}\partial_\nu\Phi\,.
\end{equation}

We consider the following action governing the dynamics of
$\Phi$ with a potential $V(\Phi)$:
\begin{equation}
S_{GSG}=-M_q^2\int d^4x\sqrt{-\det q}\sqrt{V(\Phi)}
q^{\mu\nu}\partial_\mu\Phi\partial_\nu\Phi
\,,
\end{equation}
where $M_q$ is a constant with mass dimensions.
The variation of the action with respect to $\Phi$ is
calculated as
\begin{equation}
\delta
S_{GSG}=2M_q^2\int d^4x\sqrt{-\det q}\sqrt{V(\Phi)}(\Box_q\Phi)\delta\Phi
\,,
\end{equation}
where
\begin{eqnarray}
\Box_q\Phi&\equiv&
\frac{1}{\sqrt{-\det
q}}\partial_\mu(\sqrt{-\det q}q^{\mu\nu}\partial_\nu\Phi)\,,\nonumber
\\&=&
e^{-6\Phi}V(\Phi)\left[\frac{1}{\sqrt{-\det\eta}}\partial_\mu(\sqrt{-\det\eta}\eta^{\mu\nu}
\partial_\nu\Phi)+\frac{w}{2}\frac{d}{d\Phi}\ln V(\Phi)\right]\,.
\end{eqnarray}

Novello et al.~\cite{NBMGST} have stated that the form
for $V(\Phi)$ must be chosen so as to realize
the exact Schwarzschild solution. Therefore, they employ
\begin{equation}
V(\Phi)=\frac{1}{4}{e^{2\Phi}(1-3e^{2\Phi})^2}\,.
\label{NP}
\end{equation}
We also adopt this potential in the present paper.

%%%%%%%%%%%%%%%%%%%%%%%%%%%%%%%%%%%%%%%%%%%%%%%%%%%%%%%%%%%%%%%%%%%%%%%%%%%
\subsection{Construction of hybrid model}
%%%%%%%%%%%%%%%%%%%%%%%%%%%%%%%%%%%%%%%%%%%%%%%%%%%%%%%%%%%%%%%%%%%%%%%%%%%
The following Einstein--Hilbert action is exploited in order to provide
the dynamics of the metric tensor
$g_{\mu\nu}$:
\begin{equation}
S_{GR}(g)=\frac{M_g^2}{2}\int d^4x\sqrt{-\det g}\,R_g
\,,
\label{EH}
\end{equation}
where $M_g$ is a constant with mass dimensions and $R_g$ is the
Ricci scalar constructed from the metric tensor $g$. The variation of the
action (\ref{EH}) with respect to $g$ yields 
\begin{equation}
\delta S_{GR}(g)=-\frac{M_g^2}{2}\int d^4x\sqrt{-\det
g}\left[R_g^{\mu\nu}-\frac{1}{2}R_g g^{\mu\nu}\right]\delta g_{\mu\nu}
\,,
\end{equation}
where $R_g^{\mu\nu}$ denotes the Ricci tensor constructed from $g$.

Next, we consider the mixing term of $g$ and the
effective metric $q$. For simplicity, we adopt that used in the
minimal case of the ghost-free bigravity, which is expressed as follows
\cite{HR1,HR,NO,NOS,BMMNO,BNO,NOO,BKNS,DM,KN1}:
\begin{equation}
S_{mix}(g, q)=m^2M_{0}^2\int d^4x\sqrt{-\det
g}\left[3-\tr\sqrt{g^{-1}q}+\det\sqrt{g^{-1}q}\right] \,,
\label{mix}
\end{equation}
where $m$ and $M_0$ are two constants with mass dimensions. Note that
$M_0$ is implicitly considered to be of the same order as $M_g$ and
$M_q$.   The tensor $\sqrt{g^{-1}q}$ means
$(\sqrt{g^{-1}q})^\mu{}_\rho(\sqrt{g^{-1}q})^\rho{}_\nu
=g^{\mu\rho}q_{\rho\nu}$.
The variation of (\ref{mix}) is given by
\begin{eqnarray}
\delta S_{mix}(g,q)&=&\frac{m^2M_{0}^2}{2}\int d^4x\sqrt{-\det
g}\left[g^{\mu\nu}\left(3-\tr\sqrt{g^{-1}q}\right)\right.\nonumber \\
& &\left.+\frac{1}{2}(\sqrt{g^{-1}q})^\mu{}_\rho g^{\rho\nu}
+\frac{1}{2}(\sqrt{g^{-1}q})^\nu{}_\rho g^{\rho\mu}\right]\delta
g_{\mu\nu}\nonumber \\
&+&\frac{m^2M_{0}^2}{2}\int d^4x\sqrt{-\det
g}\left[q^{\mu\nu}\det\sqrt{g^{-1}q}\right.\nonumber
\\ & &\left.-\frac{1}{2}{(\sqrt{g^{-1}q})^{-1}}^\mu{}_\rho g^{\rho\nu}
-\frac{1}{2}{(\sqrt{g^{-1}q})^{-1}}^\nu{}_\rho g^{\rho\mu}\right]\delta
q_{\mu\nu}\,.
\end{eqnarray}

Now, we define the total action for the graviton sector as the following
combination:
\begin{equation}
S=S_{GR}(g)+S_{GSG}+S_{mix}(g,q)\,.
\end{equation}
Note that a possible additional action for matter fields $S_{matter}$
will be considered later, in Sec.~\ref{sec4}. The equations of motion
derived from $S$ can be expressed as
\begin{eqnarray}
& &M_g^2\left[R_g^{\mu\nu}-\frac{1}{2}R_g g^{\mu\nu}\right]
\nonumber \\
&-&m^2M_{0}^2\left[g^{\mu\nu}\left(3-\tr\sqrt{g^{-1}q}\right)+\frac{1}{2}(\sqrt{g^{-1}q})^\mu{}_\rho
g^{\rho\nu} +\frac{1}{2}(\sqrt{g^{-1}q})^\nu{}_\rho
g^{\rho\mu}\right]=0\,,
\label{eq1}
\end{eqnarray}
and
\begin{equation}
M_q^2\sqrt{V(\Phi)}(\Box_q\Phi)
+\frac{m^2M_{0}^2}{2}\frac{\sqrt{-\det
g}}{\sqrt{-\det
q}}\left[\tau+\left(2-\frac{1}{2V}\frac{dV}{d\Phi}\right)\varepsilon+
\nabla^g_\mu
\chi^\mu\right]=0\,,
\label{eq2}
\end{equation}
where
\begin{eqnarray}
\tau&\equiv&4\det\sqrt{g^{-1}q}-\tr\sqrt{g^{-1}q}\,,
\label{ETC1}
\\
\varepsilon&\equiv&\det\sqrt{g^{-1}q}-
\frac{1}{\Omega}(\sqrt{g^{-1}q})^{-1}{}^\mu{}_\rho g^{\rho\nu}\partial_\mu
\Phi\partial_\nu\Phi\,,
\label{ETC2}
\\
\chi^\mu&\equiv&\frac{e^{-4\Phi}V-1}{\Omega}\left(-\frac{1}{2}{(\sqrt{g^{-1}q})^{-1}}^\mu{}_\rho
g^{\rho\nu} -\frac{1}{2}{(\sqrt{g^{-1}q})^{-1}}^\nu{}_\rho
g^{\rho\mu}\right.\nonumber
\\ & &\qquad\qquad\qquad
\left.+\frac{1}{\Omega}{(\sqrt{g^{-1}q})^{-1}}^\lambda{}_\rho
g^{\rho\sigma}\partial_\lambda
\Phi\partial_\sigma\Phi q^{\mu\nu}\right)
\partial_\nu\Phi\,,
\label{ETC3}
\end{eqnarray}
with
\begin{equation}
\Omega\equiv q^{\mu\nu}\partial_\mu\Phi\partial_\nu\Phi=e^{-6\Phi}Vw\,,
\quad
\nabla^g_\mu
\chi^\mu\equiv\frac{1}{\sqrt{-\det
g}}(\partial_\mu\sqrt{-\det g}\,\chi^\mu)\,.
\end{equation}
Here, we have used \cite{NBMGST}
\begin{eqnarray}
\delta q_{\mu\nu}
&=&\delta\left[e^{2\Phi}\left[\eta_{\mu\nu}-
\frac{e^{-4\Phi}V-1}{e^{-4\Phi}Vw}\partial_\mu\Phi\partial_\nu\Phi
\right]\right]\,,\nonumber \\
&=&\left[2q_{\mu\nu}+\left(4-\frac{1}{V}\frac{dV}{d\Phi}\right)
\frac{\partial_\mu\Phi\partial_\nu\Phi}{\Omega}\right]\delta\Phi
\nonumber \\
& &\quad+\frac{e^{-4\Phi}V-1}{\Omega}\left[2\frac{e^{-6\Phi}V}{\Omega}
\partial_\mu\Phi\partial_\nu\Phi\partial^\lambda\Phi\partial_\lambda\delta
\Phi-\partial_\mu\delta\Phi\partial_\nu\Phi
-\partial_\mu\Phi\partial_\nu\delta\Phi\right]\,.
\end{eqnarray}

%%%%%%%%%%%%%%%%%%%%%%%%%%%%%%%%%%%%%%%%%%%%%%%%%%%%%%%%%%%%%%%%%%%%%%%%%%%
%%%%%%%%%%%%%%%%%%%%%%%%%%%%%%%%%%%%%%%%%%%%%%%%%%%%%%%%%%%%%%%%%%%%%%%%%%%
\section{Static spherical solutions}
\label{sec3}
%%%%%%%%%%%%%%%%%%%%%%%%%%%%%%%%%%%%%%%%%%%%%%%%%%%%%%%%%%%%%%%%%%%%%%%%%%%
%%%%%%%%%%%%%%%%%%%%%%%%%%%%%%%%%%%%%%%%%%%%%%%%%%%%%%%%%%%%%%%%%%%%%%%%%%%
In this section, we consider static vacuum solutions with the spherically
symmetric ansatz.
First, we consider the flat metric in spherical coordinates
\begin{equation}
\eta_{\mu\nu}dx'^\mu dx'^\nu=-dt^2+dR^2+R^2d\Omega^2\,.
\end{equation}
where $d\Omega^2$ is the line element on a unit sphere.
Because the spherical symmetry enforces the fact that the GSG scalar field
$\Phi$ has only radial-coordinate dependence, i.e., 
$\Phi=\Phi(R)$, the effective line element $ds^2_q$ becomes
\begin{equation}
ds^2_q\equiv q'_{\mu\nu}dx'^\mu
dx'^\nu=-e^{2\Phi}dt^2+\frac{e^{6\Phi}}{V(\Phi)}dR^2+e^{2\Phi}{R^2}d\Omega^2\,.
\end{equation}
Now, converting the radial
coordinate to
$r\equiv e^\Phi R$, we find
\begin{equation}
ds^2_q=q'_{\mu\nu}dx'^\mu
dx'^\nu=q_{\mu\nu}dx^\mu
dx^\nu=-B(r)dt^2+H(r)dr^2+r^2d\Omega^2\,,
\label{stm}
\end{equation}
where
\begin{equation}
B(r)=e^{2\Phi}\,,\quad
H(r)=\frac{e^{4\Phi}}{V(\Phi)}\left(1-r\frac{d\Phi}{dr}\right)^2\,.
\label{defA}
\end{equation}

Next, we impose the bidiagonal spherically symmetric ansatz \cite{EM},
i.e., $g$ is also diagonal and assumed to be
\begin{equation}
ds^2_g=g_{\mu\nu}dx^\mu
dx^\nu=-D(r)dt^2+\frac{dr^2}{\Delta(r)}+
{r^2}\gamma(r)d\Omega^2\,.
\end{equation}

From these assumptions, the quantities defined in
(\ref{ETC1}, \ref{ETC2}, \ref{ETC3}) are expressed as
\begin{equation}
\tau=\frac{4}{\gamma}{\sqrt{\frac{HB\Delta}{D}}}-\left(
\sqrt{\frac{B}{D}}+\sqrt{H\Delta}+
\frac{2}{\sqrt{\gamma}}\right)\,,\quad
\varepsilon=\frac{1}{\gamma}{\sqrt{\frac{HB\Delta}{D}}}-\sqrt{H\Delta}\,,
\quad
\chi^\mu=0\,,
\end{equation}
and (\ref{eq2}) becomes
\begin{eqnarray}
& &\frac{|3B-1|}{2r^2\sqrt{H}}
\left({\frac{r^2 B'}{\sqrt{HB}}}\right)'+\frac{m^2M_0^2}{M_q^2}
{\gamma}{\sqrt{\frac{D}{HB\Delta}}}\left[
\frac{4}{\gamma}{\sqrt{\frac{HB\Delta}{D}}}-\left(
\sqrt{\frac{B}{D}}+\sqrt{H\Delta}+
\frac{2}{\sqrt{\gamma}}\right)\right.\nonumber \\
&
&\qquad\qquad\qquad\qquad\qquad\qquad\qquad\qquad\qquad\left.+\frac{1+3B}{1-3B}\left(
\frac{1}{\gamma}{\sqrt{\frac{HB\Delta}{D}}}-\sqrt{H\Delta}\right)
\right]=0
\,,
\label{dyq}
\end{eqnarray}
where the prime (${}'$) indicates the derivative with respect to $r$.
On the other hand, (\ref{eq1}) reads
\begin{eqnarray}
%G^0_0
& &\left(1+\frac{r\gamma'}{2\gamma}\right)
\frac{\Delta'}{r} 
+\frac{\gamma\Delta-1}{r^2\,\gamma}+ 
  \frac{\Delta}{r}\left( \frac{3\,\gamma'}{\gamma} - 
       \frac{r{\gamma'}^2}{4\,{\gamma}^2} + 
       \frac{r\gamma''}{\gamma} \right)\nonumber \\
& &-\frac{m^2M_0^2}{M_g^2}
\left[3-\left(
\sqrt{\frac{B}{D}}+\sqrt{H\Delta}+
\frac{2}{\sqrt{\gamma}}\right)+\sqrt{\frac{B}{D}}\right]=0
\,,
\label{f00}
\end{eqnarray}
\begin{eqnarray}
%G^1_1=
& &
\frac{\Delta}{r}\left(1 + \frac{r\,\gamma'}{2\,\gamma}
\right)\frac{D'}{D}
+\frac{\gamma \,\Delta -1}{r^2\,\gamma }+
\frac{\Delta }{r}
     \left( \frac{\gamma '}{\gamma } + 
       \frac{r\,{\gamma '}^2}{4\,{\gamma }^2} \right)
\nonumber \\
& &-\frac{m^2M_0^2}{M_g^2}
\left[3-\left(
\sqrt{\frac{B}{D}}+\sqrt{H\Delta}+
\frac{2}{\sqrt{\gamma}}\right)+\sqrt{H\Delta}\right]=0
\,,
\label{f11}
\end{eqnarray}
\begin{eqnarray}
%G^2_2=
& &
\frac{\Delta }{2}{\left(\frac{D''}{{D}} + \frac{D'}{r\,{D}} - 
       \frac{{D'}^2}{2\,{{D}}^2} + \frac{\gamma ''}{\gamma }+ 
       \frac{2\,\gamma '}{r\,\gamma }  - 
       \frac{{\gamma '}^2}{2\,{\gamma }^2} + 
       \frac{D'\,\gamma '}{2\,{D}\,\gamma }
        \right) }
+\frac{\Delta'}{2\,r}\left( 1 + \frac{r}{2}\left( \frac{D'}{{D}} + 
            \frac{\gamma '}{\gamma } \right)  \right)
\nonumber \\
& &-\frac{m^2M_0^2}{M_g^2}
\left[3-\left(
\sqrt{\frac{B}{D}}+\sqrt{H\Delta}+
\frac{2}{\sqrt{\gamma}}\right)+\frac{1}{\sqrt{\gamma}}\right]=0
\,.
\label{f22}
\end{eqnarray}
Note that, in the above expressions, the function $H(r)$ is defined as
\begin{equation}
H\equiv\frac{4B}{(1-3B)^2}\left(1-\frac{r}{2}\frac{B'}{B}\right)^2\,.
\label{def}
\end{equation}

From these field equations, one can find that the Schwarzschild metric is
obtained as an exact solution, i.e., 
\begin{equation}
B(r)=D(r)=\Delta(r)=1-\frac{2M_1}{r}\,,\quad \gamma(r)=1\,,
\end{equation}
where $M_1$ is an arbitrary constant.
Unfortunately, it is difficult to obtain general solutions of the
field equations explicitly, because of their severe nonlinearity.
Therefore, we perturb the metric around the Minkowski space to the first
order. Then, the field equations (\ref{dyq},
\ref{f00}, \ref{f11}, \ref{f22}) give an asymptotic solution in vacuum
having
\begin{eqnarray}
B(r)&=&1-\frac{2M_1}{r}-e^{-\mu r}\frac{M_2}{r}\,,\\
D(r)&=&1-\frac{2M_1}{r}+e^{-\mu
r}\left(\frac{\zeta M_2}{r}+O(r^{-2})\right)\,,\\
\Delta(r)&=&1-\frac{2M_1}{r}+e^{-\mu
r}\left({\mu l_0M_2}+\frac{\zeta M_2}{r}+O(r^{-2})\right)\,,\\
\gamma(r)&=&1+e^{-\mu
r}\left(\mu g_0 M_2+\frac{g_1(1+\zeta)M_2}{r}+O(r^{-2})\right)\,,
\end{eqnarray}
where $\mu$ is given by
\begin{equation}
\mu^2=\frac{2(1+\zeta)m^2M_0^2}{\zeta M_g^2+ M_q^2}\,,
\end{equation}
and $\zeta$ is a constant. The other
coefficients are determined to be
\begin{equation}
l_0=\frac{\zeta(5\zeta+2)M_g^4-2M^2_gM^2_q-M_q^4}{2M_g^2(\zeta
M_g^2+M_q^2)}\,,\quad
g_0=-\frac{\zeta M_g^2-M_q^2}{2M_g^2}\,,\quad
g_1=\frac{\zeta
M_g^2-M_q^2}{\zeta
M_g^2+M_q^2}\, .
\end{equation}
Hence, $H(r)$ is calculated as
\begin{equation}
H(r)^{-1}=1-\frac{2M_1}{r}-e^{-\mu r}\frac{M_2(1-\mu r)}{r}+\mbox{higher
orders in }M_1,M_2\mbox{ and }e^{-\mu r}\,.
\end{equation}

Interestingly, we note that 
$|q_{00}|-q_{11}^{-1}=B(r)-H(r)^{-1}=\frac{\mu M_2}{r}e^{-\mu r}
\rightarrow 0$ in the small mass limit $\mu\rightarrow 0$, up to this
order. In the same manner, we also find that 
$|g_{00}|-g_{11}^{-1}$  vanishes in the small mass limit $\mu\rightarrow
0$, up to this order. These asymptotic behaviors show a different case
from the bigravity theory with two tensor fields \cite{EM,SS}.

In this section, we have found that the static spherical solution of our
model is very similar to the GR solution.
In the next section, we consider the cosmology based on our model.

%%%%%%%%%%%%%%%%%%%%%%%%%%%%%%%%%%%%%%%%%%%%%%%%%%%%%%%%%%%%%%%%%%%%%%%%%%%
%%%%%%%%%%%%%%%%%%%%%%%%%%%%%%%%%%%%%%%%%%%%%%%%%%%%%%%%%%%%%%%%%%%%%%%%%%%
\section{Cosmological solutions}
\label{sec4}
%%%%%%%%%%%%%%%%%%%%%%%%%%%%%%%%%%%%%%%%%%%%%%%%%%%%%%%%%%%%%%%%%%%%%%%%%%%
%%%%%%%%%%%%%%%%%%%%%%%%%%%%%%%%%%%%%%%%%%%%%%%%%%%%%%%%%%%%%%%%%%%%%%%%%%%

%%%%%%%%%%%%%%%%%%%%%%%%%%%%%%%%%%%%%%%%%%%%%%%%%%%%%%%%%%%%%%%%%%%%%%%%%%%
\subsection{Cosmology with two metrics}
\label{sec41}
%%%%%%%%%%%%%%%%%%%%%%%%%%%%%%%%%%%%%%%%%%%%%%%%%%%%%%%%%%%%%%%%%%%%%%%%%%%
In this section, we attempt to study the cosmological solution for our
hybrid model. We expect new and interesting scale-factor behavior, as GSG
is known to give non-standard evolution of the scale factor
\cite{BMNT1}.

We first assume the total action as $S+S_{matter}$, where
\begin{equation}
S_{matter}=\int d^4x \sqrt{-\det g}\,{\cal L}_g(g,\varphi_g)+
\int d^4x \sqrt{-\det q}\,{\cal L}_q(q,\varphi_q)\,.
\end{equation}
This form is known as the safest and most interesting choice in bigravity
theories for cosmology
\cite{SS,AM}. It is often referred to as a ``twin matter'' model. 
Of course, this assumption involves the original Hassan-Rosen theory
\cite{HR1} for
${\cal L}_q=0$. Now, the field equations including matter are
\begin{eqnarray}
& &M_g^2\left[R_g^{\mu\nu}-\frac{1}{2}R_g g^{\mu\nu}\right]
\nonumber \\
&-&m^2M_{0}^2\left[g^{\mu\nu}\left(3-\tr\sqrt{g^{-1}q}\right)+\frac{1}{2}(\sqrt{g^{-1}q})^\mu{}_\rho
g^{\rho\nu} +\frac{1}{2}(\sqrt{g^{-1}q})^\nu{}_\rho
g^{\rho\mu}\right]=T_g^{\mu\nu}\,,
\label{eq3}
\end{eqnarray}
where
\begin{equation}
T_g^{\mu\nu}\equiv -\frac{2}{\sqrt{-\det g}}
\frac{\partial (\sqrt{-\det g}\,{\cal L}_g)}{\partial
g^{\rho\sigma}}g^{\rho\mu}g^{\sigma\nu}\,,
\end{equation}
and
\begin{eqnarray}
& &M_q^2\sqrt{V(\Phi)}(\Box_q\Phi)
+\frac{m^2M_{0}^2}{2}\frac{\sqrt{-\det
g}}{\sqrt{-\det
q}}\left[\tau+\left(2-\frac{1}{2V}\frac{dV}{d\Phi}\right)\varepsilon+
\nabla^g_\mu
\chi^\mu\right]\nonumber \\
& &=-\frac{1}{2}\left[T_q+\left(2-\frac{1}{2V}\frac{dV}{d\Phi}\right)E_q+
\nabla^q_\mu
X^\mu\right]\,,
\label{eq4}
\end{eqnarray}
where
\begin{eqnarray}
T_q&\equiv&T_q^{\mu\nu}q_{\mu\nu}\,,\quad
E_q\equiv
\frac{1}{\Omega}T_q^{\mu\nu}\partial_\mu
\Phi\partial_\nu\Phi\,,\quad
X^\mu\equiv\frac{e^{-4\Phi}V-1}{\Omega}\left(T_q^{\mu\nu}
-E q^{\mu\nu}\right)
\partial_\nu\Phi\,,
\label{ETC6}
\end{eqnarray}
with
\begin{equation}
T_q^{\mu\nu}\equiv -\frac{2}{\sqrt{-\det q}}
\frac{\partial (\sqrt{-\det q}\,{\cal L}_q)}{\partial
q^{\rho\sigma}}q^{\rho\mu}q^{\sigma\nu}\,,
\quad
\nabla^q_\mu
X^\mu\equiv\frac{1}{\sqrt{-\det
q}}(\partial_\mu\sqrt{-\det q}\,X^\mu)\,.
\end{equation}

To consider solutions of time-dependent homogeneous space,
we take the GSG scalar as a time dependent function,
$\Phi=\Phi(t')$. Then, $ds^2_q$ becomes
\begin{equation}
ds^2_q=q'_{\mu\nu}dx'^\mu
dx'^\nu=-\frac{e^{6\Phi}}{V(\Phi)}{dt'}^2+e^{2\Phi}d{\mbox{\boldmath
$x$}}^2\,.
\end{equation}
If a coordinate transformation is performed such that
$dt'=N(t)\sqrt{e^{-6\Phi}V(\Phi)}dt$, a new expression is attained:
\begin{equation}
ds^2_q=q'_{\mu\nu}dx'^\mu
dx'^\nu=q_{\mu\nu}dx^\mu
dx^\nu=-N(t)^2dt^2+e^{2\Phi(t)}d{\mbox{\boldmath $x$}}^2
=-N(t)^2dt^2+b(t)^2d{\mbox{\boldmath $x$}}^2\,,
\label{sq}
\end{equation}
where the scale factor for $q$ is defined as $b(t)\equiv e^{\Phi(t)}$.

We again consider the bidiagonal ansatz and assume that $g$
takes the form
\begin{equation}
ds^2_g=g_{\mu\nu}dx^\mu
dx^\nu
=-c(t)^2dt^2+a(t)^2d{\mbox{\boldmath $x$}}^2\,,
\label{sg}
\end{equation}
where $a(t)$ is the scale factor for $g$.
Note that the metric is the usual flat
Friedmann--Lema\^{\i}tre--Robertson--Walker metric for
$c(t)=1$.  If we set $c(t)=a(t)$, we obtain the conformal form of
this metric.

We further assume that each energy-momentum tensor is given
in the form of a perfect fluid
\begin{equation}
T_g^{\mu\nu}=(\rho_g+p_g)u_g^\mu u_g^\nu+p_g g^{\mu\nu}\,,\quad
T_q^{\mu\nu}=(\rho_q+p_q)u_q^\mu u_q^\nu+p_q q^{\mu\nu}\,,
\end{equation}
where $u_g^\mu$ and $u_q^\mu$ are the four-velocities that satisfy
$g_{\mu\nu}u_g^\mu u_g^\nu=-1$ and $q_{\mu\nu}u_q^\mu u_q^\nu=-1$,
respectively.
In the present case, the four-velocities
have the time-like component only.

Applying these ans\"atze, we find that the field equations
(\ref{eq3}, \ref{eq4}) can be rewritten as
\begin{equation}
3M_g^2\frac{\dot{a}^2}{c^2a^2}
+3m^2M_{0}^2\left(1-\frac{b}{a}\right)=\rho_g\,,
\label{cos1}
\end{equation}
\begin{equation}
M_g^2\frac{1}{c^2}\left(2\frac{\ddot{a}}{a}+\frac{\dot{a}^2}{a^2}
-2\frac{\dot{a}\dot{c}}{ac}\right)
+m^2M_{0}^2\left(3-2\frac{b}{a}-\frac{N}{c}\right)=-p_g\,,
\label{cos2}
\end{equation}
\begin{eqnarray}
& &-M_q^2\frac{b}{2N^2}|3b^2-1|\left(\frac{\ddot{b}}{b}+
2\frac{\dot{b}^2}{b^2}-\frac{\dot{b}\dot{N}}{bN}
\right)
+\frac{m^2M_{0}^2}{2}\frac{a^3c}{b^3N}
\left[\tau-\frac{3b^{2}+1}{3b^{2}-1}\varepsilon\right]\nonumber \\
&
&=-\frac{1}{2}\left[T_q-\frac{3b^{2}+1}{3b^{2}-1}E_q
\right]\,,
\label{cos3}
\end{eqnarray}
where
\begin{equation}
\tau=4\frac{b^3N}{a^3c}-3\frac{b}{a}-\frac{N}{c}\,,\quad
\varepsilon=\frac{b^3N}{a^3c}-\frac{N}{c}\,,
\label{tauepsilon}
\end{equation}
and
\begin{equation}
T_q=-\rho_q+3p_q\,,\quad
E_q=-\rho_q\,.
\end{equation}
The dot ($\dot{~}$) in the above equations denotes the derivative with
respect to the time coordinate $t$.

We further assume the conservation of the energy-momentum tensor: 
\begin{equation}
\nabla^g_\mu T_g^{\mu\nu}=\frac{1}{\sqrt{-\det g}}\partial_\mu
(\sqrt{-\det
g}\,T_g^{\mu\nu})+\Gamma(g)^\nu_{\rho\sigma}T_g^{\rho\sigma}=0\,,
\end{equation}
where
\begin{equation}
\Gamma(g)^\nu_{\rho\sigma}=\frac{1}{2}g^{\nu\lambda}(\partial_\rho 
g_{\lambda\sigma}+\partial_\sigma 
g_{\lambda\rho}-\partial_\lambda 
g_{\rho\sigma})\,.
\end{equation}
The conservation gives rise to a simple equation for the energy density
and the pressure
\begin{equation}
\dot{\rho}_g+3\frac{\dot{a}}{a}(\rho_g+p_g)=0\,.
\label{consg}
\end{equation}

Then, by applying the Bianchi identity to (\ref{eq3}), or rearranging
(\ref{cos1}), (\ref{cos2}), and (\ref{consg}), one can find a simple
relation
\begin{equation}
\frac{\dot{a}}{c}=\frac{\dot{b}}{N}\,.
\label{B}
\end{equation}

We can obtain an equation without the second derivative term 
using (\ref{cos2}), (\ref{cos3}) and (\ref{B}).
Further, using (\ref{cos1}) and (\ref{B}) again to eliminate
$\dot{a}$ and $\dot{b}$, we obtain an algebraic equation incorporating
$a$,
$b$,
$N/c$,
$\rho_g$,
$p_g$,
$\rho_q$, and $p_q$. The equation can be solved for $N/c$ and yields
{\footnotesize
\begin{equation}
\frac{N}{c}=\frac{%
\frac{\rho_g+3p_g}{3M_g^2}-\frac{6m^2M_0^2}{M_q^2}\frac{a}{b^2|3b^2-1|}
+\frac{m^2M_0^2}{M_g^2}\left(2-\frac{b}{a}\right)%
}%
{%
\frac{4a\rho_g}{3b M_g^2}-\frac{2}{a|3b^2-1|M_q^2}
\left({T_q}-\frac{3b^2+1}{3b^2-1}
E_q\right)-\frac{2m^2M_0^2}{a|3b^2-1|M_q^2}
\left[4-\frac{a^3}{b^3}-\frac{3b^2+1}{3b^2-1}
\left(1-\frac{a^3}{b^3}\right)\right]+\frac{m^2M_0^2}{M_g^2}\left(5-4\frac{a}{b}\right)}\,.
\label{N}
\end{equation}
}

We further assume 
\begin{equation}
\nabla^q_\mu T_q^{\mu\nu}=\frac{1}{\sqrt{-\det q}}\partial_\mu
(\sqrt{-\det
q}\,T_q^{\mu\nu})+\Gamma(q)^\nu_{\rho\sigma}T_q^{\rho\sigma}=0\,,
\end{equation}
where
\begin{equation}
\Gamma(q)^\nu_{\rho\sigma}=\frac{1}{2}q^{\nu\lambda}(\partial_\rho 
q_{\lambda\sigma}+\partial_\sigma 
q_{\lambda\rho}-\partial_\lambda 
q_{\rho\sigma})\,.
\end{equation}
This conservation equation implies
\begin{equation}
\dot{\rho}_q+3\frac{\dot{b}}{b}(\rho_q+p_q)=0\,.
\label{consq}
\end{equation}
Simple ans\"atze for the equations of state
\begin{equation}
p_g=\omega_g\rho_g\,,\quad p_q=\omega_q\rho_q
\end{equation}
where $\omega_g$ and $\omega_q$ are constants,
give the dependence on scale factors, i.e.,
\begin{equation}
\rho_g=\rho_{g0}\left(\frac{a_0}{a}\right)^{3(1+\omega_g)},\quad
\rho_q=\rho_{q0}\left(\frac{b_0}{b}\right)^{3(1+\omega_q)}\,,
\end{equation}
where $\rho_{g0}$, $\rho_{q0}$, $a_0$, and $b_0$ are constants. 

Using all of the ans\"atze, the $N/c$ given by (\ref{N}) can be
expressed as an function of
$a$ and $b$. Then, we can obtain the time-development of 
$a$ and $b$ by solving the differential equations
(\ref{cos1}), (\ref{B}), and (\ref{N}).

\begin{figure}[ht]
\centering
\includegraphics[height=5cm]
{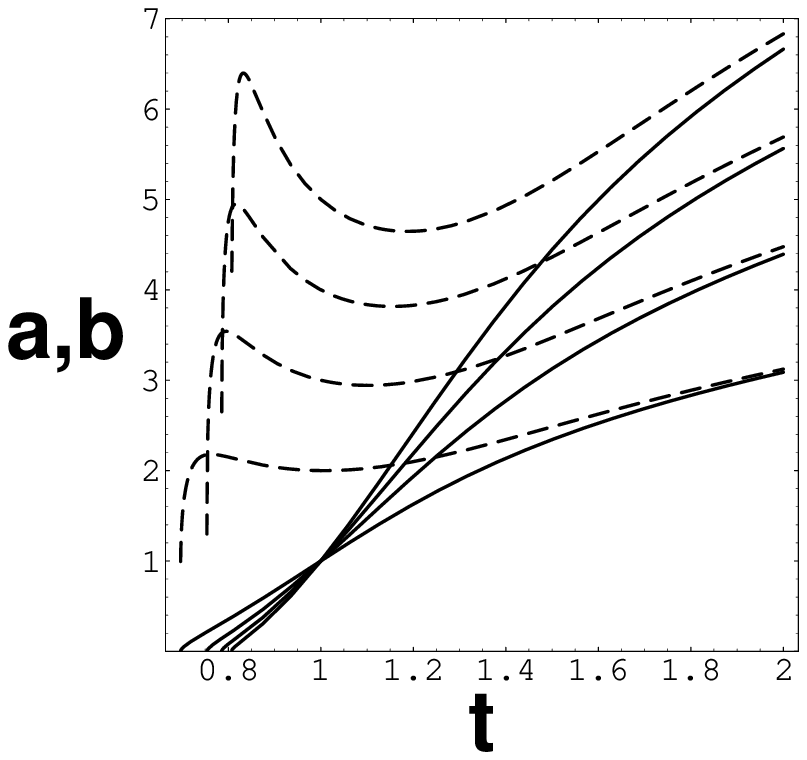}
\caption{%
Time evolution of scale factors $a$ (solid curves) and $b$ (broken curves)
with $3M_g^2m^2/\rho_{g0}=10$.
The lines correspond to $b(1)=2, 3, 4, \mbox{and } 5$. For the
parameters employed here, please see the text.}
\label{fig01}
\end{figure}

Figure~\ref{fig01} shows the results of numerical calculations.
Here, we set $M_g=M_q=M_0$ for simplicity.
We also set $c=1$ so that the parameter $t$ becomes the standard
cosmological time.
The ``initial'' value of $a(t)$ is set to $a(1)=a_0=1$.
We regard the matter as dust ($\omega_g=0$).
In these calculations, we set $\rho_q=0$ for simplicity.

\begin{figure}[ht]
\centering
\includegraphics[height=5cm]
{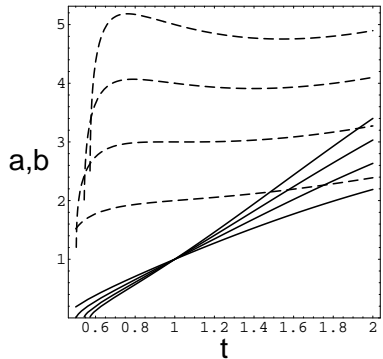}
\caption{%
Time evolution of scale factors $a$ (solid curves) and $b$ (broken curves)
with $3M_g^2m^2/\rho_{g0}=1$.
The lines correspond to $b(1)=2, 3, 4, \mbox{and } 5$. For the parameters
employed here, please see the text.}
\label{fig02}
\end{figure}

The solutions are obtained for two cases of different mass parameters
$m$: $3M_g^2m^2/\rho_{g0}=10$ (Fig.~\ref{fig01})
and $3M_g^2m^2/\rho_{g0}=1$ (Fig.~\ref{fig02}).%
\footnote{Slightly large values are used to demonstrate
accelerated expansion in the numerical results explicitly. The
acceleration can be tuned almost arbitrarily. For details, see
the discussion below (\ref{acc}).} 
As the ``initial'' value $b(1)$, we take $b(1)=2, 3, 4, \mbox{and } 5$ in
each case.

%%%%%%%%%%%%%%%%%%%%%%%%%%%
%\begin{wrapfigure}{r}%{5cm}
%\end{wrapfigure}
%%%%%%%%%%%%%%%%%%%%%%%%%%%
%%%%%%%%%%%%%%%%%%%%%%%%%%%
%\begin{wrapfigure}{r}%{5cm}
%\end{wrapfigure}
%%%%%%%%%%%%%%%%%%%%%%%%%%%

For all cases, $b(t)$ approaches $a(t)$ in the later stages. 
Thus, for the limit $t\rightarrow\infty$, the scale factors behave as the
standard Friedmann universe; this can be understood from (\ref{cos1})
and (\ref{N}). Along with the increase of $a$ and $b$, the last term of
both the numerator and denominator of $N/c$
(\ref{N}) become dominant. Hence, in later stages, we find that $a\approx
b$ and
$\dot{a}\approx \dot{b}$.

The Friedmann-like equation (\ref{cos1}) indicates that the matter is
also dominant when
$a$ is very small, as $b$ approaches $1/\sqrt{3}$ in the early stage.
We examined the case with $\rho_q\ne 0$ and found that there is no
qualitative difference in the behavior of the scale factors if there is a
comparable amount of ordinary matter ($\omega_q\ge 0$) coupled to $q$,
i.e.,
$\rho_{q0}\approx \rho_{g0}$. This is because $b$ soon becomes large,
as $N/c$ is large when
 $a$ is small: then, $\rho_q\propto
1/b^{3(1+\omega_q)}$ becomes small. 
 
\begin{figure}[ht]
\centering
\includegraphics[height=5cm]
{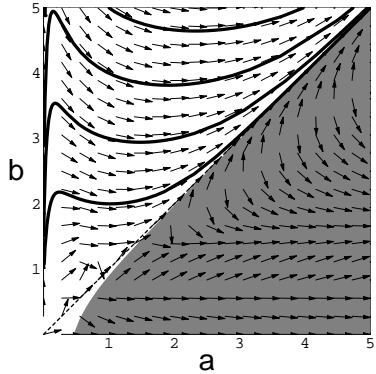}
\caption{%
Solutions plotted on $(a,b)$-plane
for $3M_g^2m^2/\rho_{g0}=10$.}
\label{fig03}
\end{figure}

\begin{figure}[ht]
\centering
\includegraphics[height=5cm]
{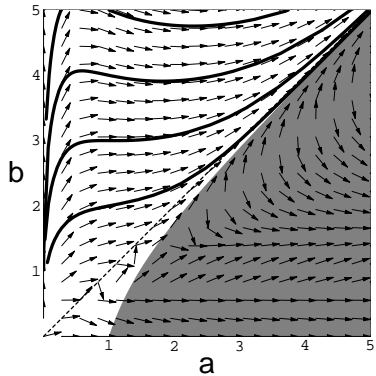}
\caption{%
Solutions plotted on $(a,b)$-plane
for $3M_g^2m^2/\rho_{g0}=1$.}
\label{fig04}
\end{figure}

The relative evolution of the two scale factors can be clearly seen 
if the solutions are plotted on an $(a,b)$-plane as shown in
Figs.~\ref{fig03} and \ref{fig04}. The arrows in the figures indicate the
normalized vector
\begin{equation}
\frac{1}{\sqrt{\dot{a}^2+\dot{b}^2}}(\dot{a},
\dot{b})=\frac{1}{\sqrt{1+N^2/c^2}}(1, N/c)\,,
\end{equation}
at each point.
In these figures, the shaded regions indicate that the right hand side of
(\ref{cos1}) becomes negative.
%%%%%%%%%%%%%%%%%%%%%%%%%%%
%\begin{wrapfigure}{r}%{5cm}
%\end{wrapfigure}
%%%%%%%%%%%%%%%%%%%%%%%%%%%
%%%%%%%%%%%%%%%%%%%%%%%%%%%
%\begin{wrapfigure}{r}%{5cm}
%\end{wrapfigure}
%%%%%%%%%%%%%%%%%%%%%%%%%%%

Returning to Figs.~\ref{fig01} and \ref{fig02}, we find that 
accelerated expansion occurs for a relatively large $b(1)$.
From (\ref{cos1}) and (\ref{cos2}), one arrives at the equation
\begin{equation}
2\frac{1}{ca}\left(\frac{\dot{a}}{c}\right)^{\cdot}=-\frac{1}{3M_g^2}(\rho_g+3p_g)-\frac{m^2M_0^2}{M_g^2}
\left(2-\frac{b}{a}+\frac{N}{c}\right)\,.
\label{acc}
\end{equation}
If we take $c=1$ here, in other words, $t$ is the standard cosmological
time in the system described by $g$, the left hand side of
(\ref{acc}) reads $2\ddot{a}/a$.
Thus, we can confirm that accelerated expansion is only possible if
$b/a-N/c>2$.
Because the value of $N/c$ is negative or almost zero at $t=1$, 
cosmic acceleration is feasible for a large value of
$b(1)/a(1)$ and for a large value of $3M_g^2m^2/\rho_{g0}$, which is
confirmed by the numerical calculations.

A subtle point to note is that $|q_{00}|=N^2$ vanishes at certain points
for sufficiently large $b(1)$ and $m^2$: in other words, there are
determinant singularities
\cite{GHW}. As we hold that $q$ is not a genuine metric, no
problem exists especially in the case of
${\cal L}_q=0$. The degenerate metric, however, may induce field
theoretical problems if the matter field is coupled to
$q$, i.e., ${\cal L}_q\ne 0$.

In this subsection, we have found that the $a$ of the
physical $g$ can exhibit accelerated expansion if
 $m$ and the initial value of $b/a$ are sufficiently large,
even if there is no exotic matter.
It is worth noting that the expansion decelerates in the early stages.
The acceleration occurs subsequently and ends in the later stages.

%%%%%%%%%%%%%%%%%%%%%%%%%%%%%%%%%%%%%%%%%%%%%%%%%%%%%%%%%%%%%%%%%%%%%%%%%%%
\subsection{Cosmology with composite metric}
\label{sec4B}
%%%%%%%%%%%%%%%%%%%%%%%%%%%%%%%%%%%%%%%%%%%%%%%%%%%%%%%%%%%%%%%%%%%%%%%%%%%

Next, we consider a model with a ``composite'' metric, similar to the 
bigravity models proposed in
\cite{dRHR,SM}. The action for matter is now assumed to be described with
the composite metric
\begin{equation}
G_{\mu\nu}=\alpha^2 g_{\mu\nu}+2\alpha\beta g_{\mu\rho}
(\sqrt{g^{-1}q})^\rho{}_\nu+\beta^2 q_{\mu\nu}\,,
\end{equation}
where $\alpha$ and $\beta$ are constants.
The composite line element for cosmology is given by
\begin{equation}
ds_G^2=G_{\mu\nu}dx^\mu dx^\nu=-(\alpha c+\beta N)^2dt^2+(\alpha a+\beta
b)^2 d{\mbox{\boldmath
$x$}}^2\equiv -{C}^2dt^2+A^2 d{\mbox{\boldmath
$x$}}^2\,,
\end{equation}
where we use the same symbols for the components of the two metrics as in
the previous subsection. The energy-momentum tensor of the matter field is
understood to be conserved with respect to the description when
the composite metric is employed, i.e.,
\begin{equation}
\nabla^G_\mu T_G^{\mu\nu}=\frac{1}{\sqrt{-\det G}}\partial_\mu
(\sqrt{-\det
G}\,T_G^{\mu\nu})+\Gamma(G)^\nu_{\rho\sigma}T_G^{\rho\sigma}=0\,,
\end{equation}
and we further assume the perfect fluid form for the isotropic
and homogeneous universe to be
\begin{equation}
T_G^{\mu\nu}=(\rho_G+p_G)u_G^\mu u_G^\nu+p_G
G^{\mu\nu}\,,
\end{equation} where $u_G^\mu$ satisfies
$G_{\mu\nu}u_G^\mu u_G^\nu=-1$.

The field equations are found using the treatment in \cite{SM} and
following a similar calculation to previously, yielding
\begin{equation}
3M_g^2\frac{\dot{a}^2}{c^2a^2}
+3m^2M_{0}^2\left(1-\frac{b}{a}\right)=\alpha\frac{A^3}{a^3}\rho_G\,,
\label{co1}
\end{equation}
\begin{equation}
M_g^2\frac{1}{c^2}\left(2\frac{\ddot{a}}{a}+\frac{\dot{a}^2}{a^2}
-2\frac{\dot{a}\dot{c}}{ac}\right)
+m^2M_{0}^2\left(3-2\frac{b}{a}-\frac{N}{c}\right)
=-\alpha\frac{{C}A^2}{ca^2}p_G\,,
\label{co2}
\end{equation}
\begin{eqnarray}
& &-M_q^2\frac{b}{2N^2}|3b^2-1|\left(\frac{\ddot{b}}{b}+
2\frac{\dot{b}^2}{b^2}-\frac{\dot{b}\dot{N}}{bN}
\right)
+\frac{m^2M_{0}^2}{2}\frac{a^3c}{b^3N}
\left[\tau-\frac{3b^{2}+1}{3b^{2}-1}\varepsilon\right]\nonumber \\
&
&=-\frac{1}{2}\frac{A^3}{b^3}\beta\left[
\frac{2}{3b^{2}-1}\rho_G+3\frac{Cb}{NA}p_G
\right]\,,
\label{co3}
\end{eqnarray}
with the definition given in (\ref{tauepsilon}).
Then, the Bianchi identity yields the same relation as (\ref{B}).%
\footnote{To be precise, the identity leads to
$\left(m^2M_0^2-\alpha\beta\frac{A^2}{a^2}p_G\right)\left(\dot{b}
-\frac{N}{c}\dot{a}\right)=0$.}

We can solve all the above equations to obtain
\begin{equation}
\frac{N}{c}=\frac{\cal N}{\cal C}\,,
\label{r}
\end{equation}
where
\begin{eqnarray}
{\cal N}&=&
\frac{A^3\alpha}{3a^3M_g^2}\left(\rho_G+\frac{3\alpha a}{A}p_G\right)
+\frac{2 A^2\alpha\beta}{ab^2|3b^2-1|M_q^2}p_G\nonumber \\
& &-\frac{6m^2M_0^2}{M_q^2}\frac{a}{b^2|3b^2-1|}
+\frac{m^2M_0^2}{M_g^2}\left(2-\frac{b}{a}\right)\,,
\label{r1}
\end{eqnarray}
and
\begin{eqnarray}
{\cal C}&=&\frac{4A^3\alpha}{3ba^2
M_g^2}\left(\rho_G-
\frac{3\beta b}{4A}p_G\right)-\frac{2A^3\beta}{ab^3|3b^2-1|M_q^2}
\left(\frac{2}{3b^2-1}\rho_G+\frac{3\beta b}{A}p_G\right)\nonumber \\
& &
-\frac{2m^2M_0^2}{a|3b^2-1|M_q^2}
\left[4-\frac{a^3}{b^3}-\frac{3b^2+1}{3b^2-1}
\left(1-\frac{a^3}{b^3}\right)\right]+\frac{m^2M_0^2}{M_g^2}
\left(5-4\frac{a}{b}\right)\,.
\label{r2}
\end{eqnarray}

Because $N/c=\dot{b}/\dot{a}$, (\ref{co1}) and 
(\ref{r}, \ref{r1}, \ref{r2}) can express the development of the scale
factors, if the equation of state for matter is given.
Note that $N/c\rightarrow 1$ if $\rho_G, p_G\rightarrow 0$ and
$b\rightarrow 0$, as in the case examined in the previous subsection.

Here, we again
take a simple assumption for the equation of state
\begin{equation}
p_G=\omega_G\rho_G\,,
\end{equation}
where $\omega_G$ is a constant. Then, the dependence on the scale factor
$A$ is 
\begin{equation}
\rho_G=\rho_{G0}\left(\frac{A_0}{A}\right)^{3(1+\omega_G)}\,,
\end{equation}
where $\rho_{G0}$ and $A_0$ are constants.

Note that the following relation holds:
\begin{equation}
\frac{C}{c}=\frac{\alpha
c+\beta
N}{c}=\alpha+\beta\frac{N}{c}=\alpha+\beta\frac{\dot{b}}{\dot{a}}
=\frac{\alpha
\dot{a}+\beta\dot{b}}{\dot{a}}=\frac{\dot{A}}{\dot{a}}\,.
\label{es}
\end{equation}
If we choose a new cosmological time $T$, which satisfies $dT=Cdt$,
this relation is no more than $\frac{dA}{dT}=\frac{\dot{a}}{c}$.
Thus, if $\dot{a}/c>0$, $A$ also increases with $T$.

From (\ref{co1}) and (\ref{co2}), we find the second-order
differential equation
\begin{equation}
2\frac{1}{ca}\left(\frac{\dot{a}}{c}\right)^{\cdot}=-\frac{\alpha}{3M_g^2}\left(\frac{A^3}{a^3}
\rho_G+3\frac{A^2}{a^2}p_G\right)-\frac{m^2M_0^2}{M_g^2}
\left(2-\frac{b}{a}+\frac{N}{c}\right)\,.
\end{equation}
Using $T$, this equation becomes
\begin{equation}
2\frac{C}{ca}\frac{d^2{A}}{dT^2}
=-\frac{\alpha}{3M_g^2}\left(\frac{A^3}{a^3}
\rho_G+3\frac{A^2}{a^2}p_G\right)-\frac{m^2M_0^2}{M_g^2}
\left(2-\frac{b}{a}+\frac{N}{c}\right)\,.
\end{equation}
Provided $C/c$ is positive,
we expect that accelerated expansion of $A(T)$ is only possible if
$b/a-N/c>2$, as in the case treated in the previous subsection.

\begin{figure}[ht]
\centering
\includegraphics[height=5cm]
{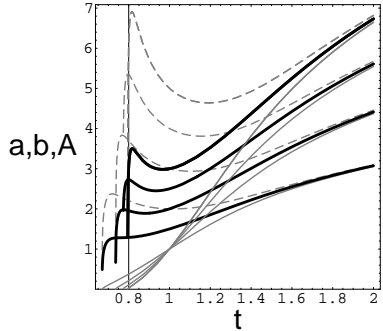}
\caption{%
Time evolution of scale factors $A=0.5 a+0.5 b$ (solid curves),
$a$ (gray solid curves), and
$b$ (gray broken curves). The lines correspond to $b(1)=2, 3, 4,
\mbox{and } 5$. For the parameters employed here, please see the text.}
\label{fig05}
\end{figure}

Figure~\ref{fig05} shows the results of numerical calculations for the
composite metric model. Here, we set $M_g=M_q=M_0$ for simplicity.
We also set $c=1$ so as to make the parameter $t$ the standard
cosmological time. The parameters $(\alpha, \beta)$ are taken to be
$(0.5, 0.5)$, $m$ is chosen to satisfy
$3M_g^2m^2/\rho_{G0}=10$, and the ``initial'' value of the scale factor
$a(t)$ is set to $a(1)=a_0=1$. We consider dust matter ($\omega_G=0$).
As the ``initial'' value $b(1)$, we take $b(1)=2, 3, 4, \mbox{and }
5$.

%%%%%%%%%%%%%%%%%%%%%%%%%%%
%\begin{wrapfigure}{r}%{5cm}
%\end{wrapfigure}
%%%%%%%%%%%%%%%%%%%%%%%%%%%

The behaviors of $a$ and $b$ are similar to the case treated in the
previous subsection.
This time, however, the ``physical'' and unique scale factor is $A$.
One can see the novel behavior of $A(t)$ from Fig.~\ref{fig05};
however, we must bear in mind that $t$ is not the most
suitable  cosmological time.

\begin{figure}[ht]
\centering
\includegraphics[height=5cm]
{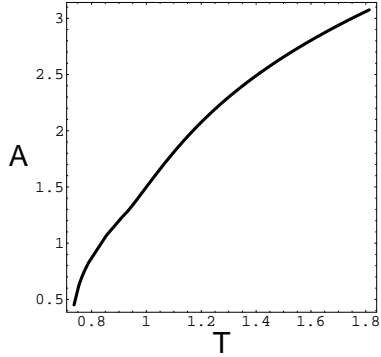}
\caption{%
Plot of $A(T)$ for $b(1)=2$ and $\alpha=\beta=0.5$
($T=1$ at $t=1$).}
\label{fig06}
\end{figure}

We plot $A(T)$ for $b(1)=2$ against $T$ in
Fig.~\ref{fig06}. For small and large $T$, the solution resembles
that of the Friedmann universe. $A$ only seems to accelerate in the
intermediate era.
 
%%%%%%%%%%%%%%%%%%%%%%%%%%%
%\begin{wrapfigure}{r}%{5cm}
%\end{wrapfigure}
%%%%%%%%%%%%%%%%%%%%%%%%%%%

\begin{figure}[ht]
\centering
\includegraphics[height=5cm]
{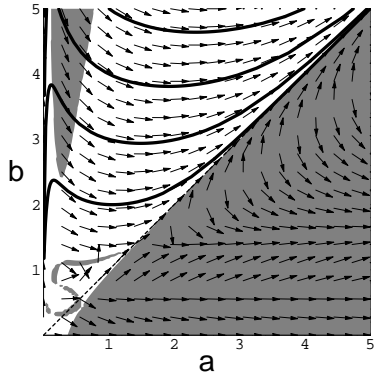}
\caption{%
Solutions plotted on $(a,b)$-plane for $\alpha=\beta=0.5$.}
\label{fig07}
\end{figure}

Figure~\ref{fig07} shows a plot of the solutions on the $(a,b)$-plane.
In this figure, the shaded region indicates where the right hand side of
(\ref{cos1}) or $C$ becomes negative.
For a sufficiently large $b(1)$, we find that the solution passes the
point where $C=0$ (and $\frac{dA}{dt}=0$ at the same time, for
(\ref{es})).
The degenerate metric with $G_{00}=0$ is too curious to assign a 
physical meaning.

%%%%%%%%%%%%%%%%%%%%%%%%%%%
%\begin{wrapfigure}{r}%{5cm}
%\end{wrapfigure}
%%%%%%%%%%%%%%%%%%%%%%%%%%%

\begin{figure}[ht]
\centering
\includegraphics[height=5cm]
{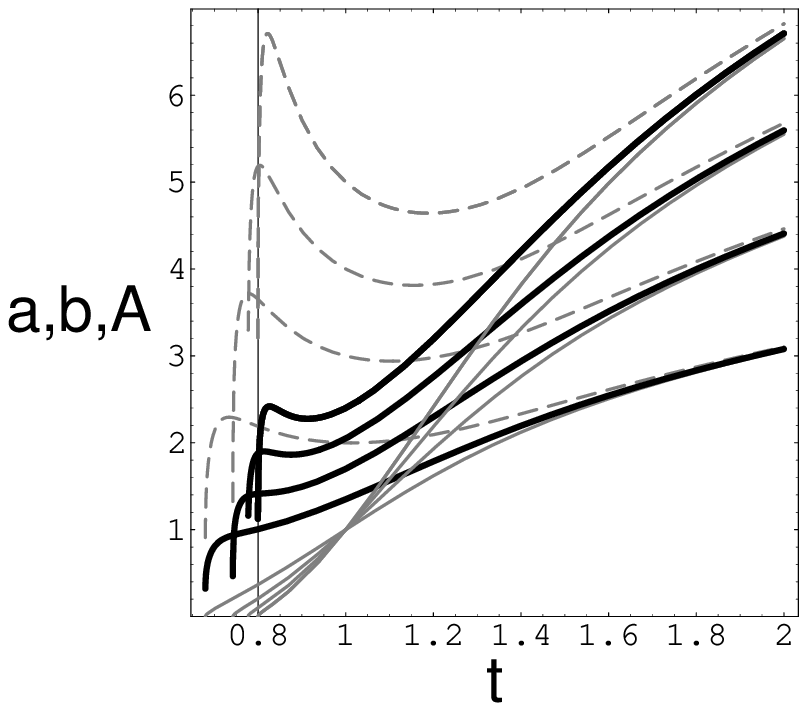}
\caption{%
Solutions plotted on $(a,b)$-plane for $\alpha=0.65$ and
$\beta=0.35$.}
\label{fig08}
\end{figure}

\begin{figure}[ht]
\centering
\includegraphics[height=5cm]
{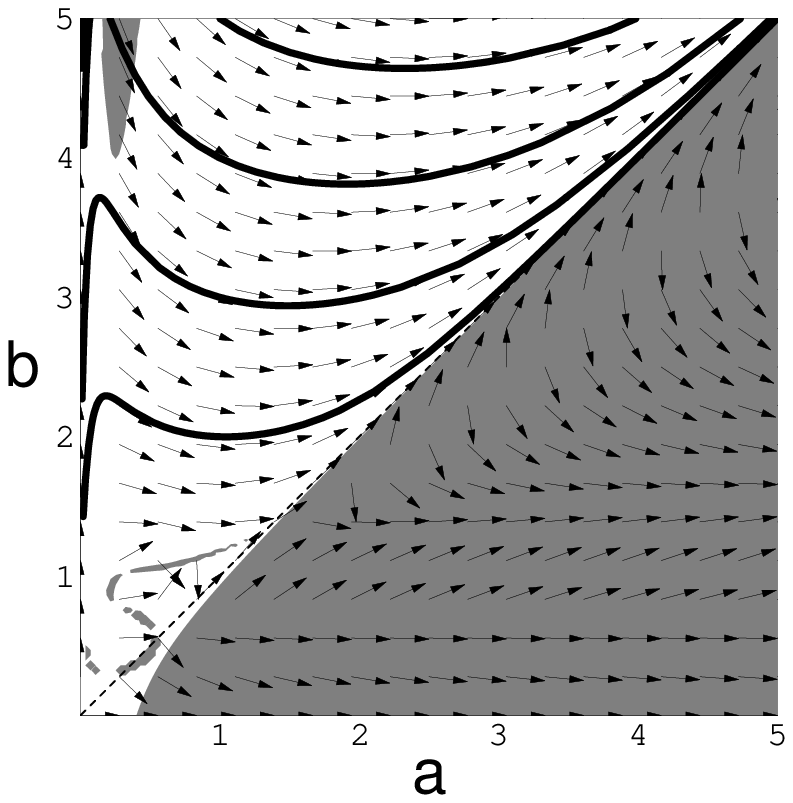}
\caption{%
Solutions plotted on $(a,b)$-plane for $\alpha=0.65$ and
$\beta=0.35$.}
\label{fig09}
\end{figure}

Thus, the initial value $b(1)$ cannot be large, for instance, $b(1)$
must be less than $\approx 2$ in the case of $\alpha=\beta=0.5$. 
If the value of $\alpha/\beta$ is larger, larger $b(1)$ is allowed.
This fact can be seen from Figs.~\ref{fig08} and \ref{fig09}, in which
the solutions for
$\alpha=0.65$ and $\beta=0.35$ are plotted. 

%%%%%%%%%%%%%%%%%%%%%%%%%%%
%\begin{wrapfigure}{r}%{5cm}
%\end{wrapfigure}
%%%%%%%%%%%%%%%%%%%%%%%%%%%

%%%%%%%%%%%%%%%%%%%%%%%%%%%
%\begin{wrapfigure}{r}%{5cm}
%\end{wrapfigure}
%%%%%%%%%%%%%%%%%%%%%%%%%%%

\begin{figure}[ht]
\centering
\includegraphics[height=5cm]
{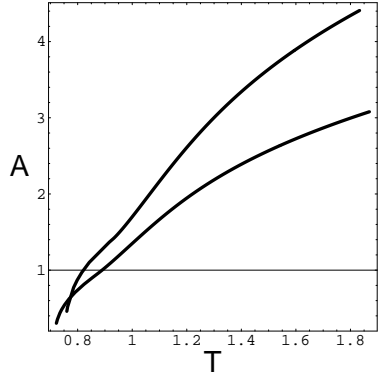}
\caption{%
Plot of $A(T)$ for $b(1)=2$ (lower curve) and $b(1)=3$ (upper curve)
for
$\alpha=0.65$ and
$\beta=0.35$ ($T=1$ at $t=1$).}
\label{fig10}
\end{figure}

The evolution of $A(T)$ is plotted in Fig.~\ref{fig10} for
$\alpha=0.65$ and
$\beta=0.35$.
A larger initial value of $b/a$ yields larger acceleration
in the permitted parameter range.

%%%%%%%%%%%%%%%%%%%%%%%%%%%
%\begin{wrapfigure}{r}%{5cm}
%\end{wrapfigure}
%%%%%%%%%%%%%%%%%%%%%%%%%%%

In this subsection, we have found that the ``physical'' scale factor
$A=\alpha a+\beta b$ can exhibit accelerated expansion if 
$m$ and the initial value of $b/a$ are sufficiently large.
The parameters are restricted by the condition that the ``physical''
metric should be non-degenerate (i.e., $G_{00}$ does not vanish). The
expansion shows successive 
deceleration, acceleration, and deceleration.

%%%%%%%%%%%%%%%%%%%%%%%%%%%%%%%%%%%%%%%%%%%%%%%%%%%%%%%%%%%%%%%%%
%%%%%%%%%%%%%%%%%%%%%%%%%%%%%%%%%%%%%%%%%%%%%%%%%%%%%%%%%%%%%%%%%
\section{Summary and prospects}
\label{summary}
%%%%%%%%%%%%%%%%%%%%%%%%%%%%%%%%%%%%%%%%%%%%%%%%%%%%%%%%%%%%%%%%%
%%%%%%%%%%%%%%%%%%%%%%%%%%%%%%%%%%%%%%%%%%%%%%%%%%%%%%%%%%%%%%%%%

In this paper, we have presented a GR--GSG hybrid model of gravity. We
have shown that
the exact Schwarzschild solution is produced
and the accelerating phase of the universe
is obtained without the cosmological constant in this model.
In this paper, we have shown only qualitative analyses,
because there are many tunable parameters of our model, such as
$M_q/M_g$, $\rho_{q0}$, and $\omega_g (\omega_q)$, as well as $m^2$,
$b(1)$, and $a(1)$. Moreover, we can assume a general mixing of $g$
and $q$ in $S_{mix}(g, q)$ other than the minimal choice considered in the
present paper. Further research should be followed in future.

Unfortunately, our model does not provide a mechanism for inflation.
However, because the early phase of the universe in our model resembles
the Friedmann universe, incorporation of the inflation dynamics can be
naturally introduced in the very early phase. The aspect
of inflation in the GR--GSG hybrid model is an important subject for
future study.

The quantum cosmology of our GR--GSG hybrid model is another very
interesting subject, as the evolution of scale factors is naively
dependent on the initial conditions. In particular, our classical model
cannot avoid the singularity problem, unfortunately.
Quantum cosmological approaches to the problem of singularities 
are common topics of study to which bimetric theories are
applied \cite{DM}.

In future, we hope to investigate many aspects of GR--GSG hybrid
gravity, such as compact objects, instability problems (including
initial fluctuations%
\footnote{If twin matter exists in our model (${\cal L}_q\ne 0$), the
primordial fluctuations may exhibit novel behavior because of the rapid
expansion of
$b(t)$. However, the problem of a degenerate effective metric arises for
some parameter choices.}), and anisotropic solutions in the model, as
well as the above-mentioned subjects. Through these investigations,  we
will find the phenomenological limit of the theory and obtain the ability
to construct a more realistic model based on the present model.

\acknowledgments
%%%%%%%%%%%%%%%%%%%%%%%%%%%%%%%%%%%%%%%%%%%%%%%%%%%%%%%%%%%%%%%%%%%%%%%%%%%
%Acknowledgements
%%%%%%%%%%%%%%%%%%%%%%%%%%%%%%%%%%%%%%%%%%%%%%%%%%%%%%%%%%%%%%%%%%%%%%%%%%%
%\begin{acknowledgments}
This study is supported in part by a Grant-in-Aid from the Nikaido
Research  Fund.
%the organizers of JGRG21, where our
%partial result %({\tt [arXiv:10mm.xxxx]}) 
%was presented. %for elucidating comments.
%This study is supported in part by the Grant-in-Aid of Nikaido Research 
%Fund.
%\end{acknowledgments}

%%%%%%%%%%%%%%%%%%%%%%%%%%%%%%%%%%%%%%%%%%%%%%%%%%%%%%%%%%%%%%%%%%%%%%%%%%%
%%%%%%%%%%%%%%%%%%%%%%%%%%%%%%%%%%%%%%%%%%%%%%%%%%%%%%%%%%%%%%%%%%%%%%%%%%%
%%%%%%%%%%%%%%%%%%%%%%%%%%%%%%%%%%%%%%%%%%%%%%%%%%%%%%%%%%%%%%%%%%%%%%%%%%%
%%%%%%%%%%%%%%%%%%%%%%%%%%%%%%%%%%%%%%%%%%%%%%%%%%%%%%%%%%%%%%%%%%%%%%%%%%%

%%%%%%%%%%%%%%%%%%%%%%%%%%%%%%%%%%%%%%%%%%%%%%%%%%%%%%%%%%%%%%%%%%%%%%%%%%%
%thebibliography
%%%%%%%%%%%%%%%%%%%%%%%%%%%%%%%%%%%%%%%%%%%%%%%%%%%%%%%%%%%%%%%%%%%%%%%%%%%
%\bibliographystyle{apsrev}
\bibliographystyle{apsrev4-1}
%\bibliography{}

%%%%%%%%%%%%%%%%%%%%%%%%%%%%%%%%%%%%%%%%%%%%%
%%%%%%%%%%%%%%%%%%%%%%%%%%%%%%%%%%%%%%%%%%%%%
%%%%%%%%%%%%%%%%%%%%%%%%%%%%%%%%%%%%%%%%%%%%%
%%%%%%%%%%%%%%%%%%%%%%%%%%%%%%%%%%%%%%%%%%%%%
%%%%%%%%%%%%%%%%%%%%%%%%%%%%%%%%%%%%%%%%%%%%%
%%%%%%%%%%%%%%%%%%%%%%%%%%%%%%%%%%%%%%%%%%%%%
%%%%%%%%%%%%%%%%%%%%%%%%%%%%%%%%%%%%%%%%%%%%%

%%%%%%%%%%%%%%%%%%%%%%%%%%%%%%%%%%%%%%%%%%%%%%%%%%%%%%%%%%%%%%%%%%%%%%%%%%%
%%%%%%%%%%%%%%%%%%%%%%%%%%%%%%%%%%%%%%%%%%%%%%%%%%%%%%%%%%%%%%%%%%%%%%%%%%%
%%%%%%%%%%%%%%%%%%%%%%%%%%%%%%%%%%%%%%%%%%%%%%%%%%%%%%%%%%%%%%%%%%%%%%%%%%%
%%%%%%%%%%%%%%%%%%%%%%%%%%%%%%%%%%%%%%%%%%%%%%%%%%%%%%%%%%%%%%%%%%%%%%%%%%%
%%%%%%%%%%%%%%%%%%%%%%%%%%%%%%%%%%%%%%%%%%%%%%%%%%%%%%%%%%%%%%%%%%%%%%%%%%%

\end{document}